\documentclass[12pt]{article}

\usepackage{epsfig}

\usepackage{graphicx}

\def\square{\kern1pt\vbox{\hrule height 1.2pt
\hbox{\vrule width 1.2pt\hskip 3pt
\vbox{\vskip 6pt}\hskip 3pt\vrule width 0.6pt}
\hrule height 0.6pt}\kern1pt}
\def\ltwid{\mathrel{\raise.3ex\hbox{$<$\kern-.75em\lower1ex\hbox{$\sim$}}}}

\begin{document}

\begin{titlepage}
\begin{flushright}
CCTP-2012-06 \\
UFIFT-QG-12-03
\end{flushright}

\vspace{0.5cm}

\begin{center}
\bf{Quantum Gravity and Inflation}
\end{center}

\vspace{0.3cm}

\begin{center}
Maria G. Romania$^{\dagger}$ and N. C. Tsamis$^{\ddagger}$
\end{center}
\begin{center}
\it{Department of Physics, University of Crete \\
GR-710 03 Heraklion, HELLAS.}
\end{center}

\vspace{0.2cm}

\begin{center}
R. P. Woodard$^{\ast}$
\end{center}
\begin{center}
\it{Department of Physics, University of Florida \\
Gainesville, FL 32611, UNITED STATES.}
\end{center}

\vspace{0.3cm}

\begin{center}
ABSTRACT
\end{center}
\hspace{0.3cm} We review some perturbative results obtained
in quantum gravity in an accelerating cosmological background.
We then describe a class of non-local, purely gravitational
models which have the correct structure to reproduce the
leading infrared logarithms of quantum gravitational
back-reaction during the inflationary regime. These models
end inflation in a distinctive phase of oscillations with
slight and short violations of the weak energy condition
and should, when coupled to matter, lead to rapid reheating.
By elaborating this class of models we exhibit one that has
the same behaviour during inflation, goes quiescent until
the onset of matter domination, and induces a small, positive
cosmological constant of about the right size thereafter.
We also briefly comment on the primordial density perturbations
that this class of models predict.

\vspace{0.3cm}

\begin{flushleft}
\end{flushleft}

\vspace{0.1cm}

\begin{flushleft}
$^{\dagger}$ e-mail: romania@physics.uoc.gr \\
$^{\ddagger}$ e-mail: tsamis@physics.uoc.gr \\
$^{\ast}$ e-mail: woodard@phys.ufl.edu
\end{flushleft}

\end{titlepage}

\section{Introduction}

${\bullet \;}$ {\bf FRW Cosmology and Inflation:}
\\ [2pt]
On scales larger than about $100 Mpc$ the universe is well 
described by the $FRW$ geometry:
\begin{equation}
ds^2 \; = \;
- dt^2 \, + \, a^2(t) \, d{\bf x} \cdot d{\bf x}
\;\; . \label{frw}
\end{equation}
The time variation of the scale factor $a(t)$ gives the
instantaneous values of the {\it Hubble parameter}
$H(t)$  and the {\it deceleration parameter} $q(t)$ :
\begin{eqnarray}
H(t) &\!\! \equiv \!\!&
\frac{\dot{a}(t)}{a(t)}
\; = \; \frac{d}{dt} \ln a(t)
\;\; , \label{H} \\
q(t) &\!\! \equiv \!\!&
- \frac{\dot{a}(t) \; \ddot{a}(t)}{\dot{a}^2(t)}
\; = \;
-1 - \frac{\dot{H}(t)}{H^2(t)}
\; \equiv \;
-1 + \epsilon(t)
\;\; . \label{q}
\end{eqnarray}
Their current values are:
$H_{now} \simeq (73.8 \pm 2.4) km/sec \, Mpc \simeq 
2.4 \times 10^{-18} Hz$ \cite{Riess} and 
$\epsilon_{now} \simeq 0.33 \pm 0.13$ \cite{Hicken,Komatsu}.

There is overwhelming evidence that the history of the universe
included a period of accelerated expansion known as inflation and
defined by $H > 0$ with $\epsilon < 1$ \cite{Linde,Slava}. This
expansion occured very early -- $t \sim 10^{-33}sec$ -- and the
latest data \cite{Komatsu, Keisler}, plus the assumption of single
scalar inflation imply: $H_I \ltwid 1.7 \times 10^{38} Hz$ with
$\epsilon_I \ltwid 0.011$ \cite{KOW}.
\\ [4pt]
${\bullet \;}$ {\bf The Horizon Problem:}
\\ [2pt]
The strongest evidence in favour of primordial inflation is the 
fact that we can detect epochs of cosmological history during 
which the observable universe was in thermal equilibrium. Without 
an early phase of primordial acceleration there is no way such 
distant regions can even have exchanged a single photon, much less
interacted strongly enough to have equilibrated. To understand why,
let us use the fact that photons travel on paths with zero invariant
interval to calculate the size of our horizon:
\begin{equation}
ds^2 \, = \,
- dt^2 \, + \, a^2(t) \, dr^2 \, = \, 0
\quad \Longrightarrow \quad
dr \, = \, \frac{dt}{a(t)}
\;\; . \label{frwlight}
\end{equation}
Now consider some past time $t_{past}$, and compare the coordinate
distance $R_{past}$ we can observe at $t_{now}$ with the coordinate
radius of light which propagated from the beginning of the universe
at $t_{initial}$ to $t_{past}$:
\begin{eqnarray}
R_{past} &\!\! = \!\!&
\int \frac{dt}{a(t)}
\quad , \quad
{\rm for} \;\; t_{past} \; < \; t \; < \; t_{now}
\;\; , \label{Rpast} \\
R_{future} &\!\! = \!\!&
\int \frac{dt}{a(t)}
\quad , \quad
{\rm for} \;\; t_{initial} \; < \; t \; < t_{past}
\;\; . \label{Rfuture}
\end{eqnarray}
For the universe at $t_{past}$ to have reached thermal equilibrium
by causal processes requires:
\begin{equation}
\left( \frac{R_{past}}{R_{future}} \right)^2 \; \leq \; 1
\;\; . \label{causality}
\end{equation}

Suppose the universe expanded with constant $\epsilon \equiv 
-{\dot H} H^{-2}$:
\begin{eqnarray}
{\rm constant} \; \epsilon
& \Longrightarrow &
H = \frac{1}{\epsilon t}
\quad \Longrightarrow \quad
a \sim t^{\frac{1}{\epsilon}}
\label{constepsilon1} \\
& \Longrightarrow &
\int \frac{dt}{a(t)} =
\frac{1}{(\epsilon -1) H a}
\;\; . \label{constepsilon2}
\end{eqnarray}
If the universe was decelerating throughout its existence the upper
limit of (\ref{constepsilon2}) dominates over the lower one:
\begin{eqnarray}
\epsilon > 1
& \Longrightarrow &
Ha \; \sim \; t^{1 - \frac{1}{\epsilon}} \;\; {\rm falls}
\label{dec1} \\
& \Longrightarrow &
R_{past} \; \sim \; \frac{1}{(\epsilon -1) H a} \,
\Big|_{now}
\;\; , \quad
R_{future} \; \sim \; \frac{1}{(\epsilon -1) H a} \,
\Big|_{past}
\qquad \label{dec2} \\
& \Longrightarrow & R_{future} \; \ll \; R_{past} \;\; .
\label{dec3}
\end{eqnarray}
For instance, at recombination -- when the universe is observed 
to be in thermal equilibrium to one part in $10^5$! -- and at 
nucleosynthesis, respectively:
\begin{equation}
\left( \frac{R_{past}}{R_{future}} \right)^2 \; \sim \; 2000 \;\;
{\rm and} \;\; 10^9 \;\; . \label{reconucleo}
\end{equation}
Hence the observed equilibrium during these epochs could not have
come about by causal processes; it would have had to be an
accidental feature of the way the universe began. No one knows how
the universe began, but assuming it began in a very high degree of
thermal equilibrium seems problematic. This sort of unsatisfactory
conclusion can be avoided if we assume the universe went through a
phase of acceleration before $t_{past}$. In that case the integral
(\ref{constepsilon2}) is dominated by its lower limit and we can
make the past light-cone arbitrarily large by assuming $t_{initial}$
is close to zero:
\begin{eqnarray}
\epsilon < 1
& \Longrightarrow &
Ha \; \sim \; t^{1 - \frac{1}{\epsilon}} \;\; {\rm grows}
\label{acc1} \\
& \Longrightarrow &
R_{future} \, \sim \, \frac{1}{(\epsilon -1) H a} \,
\Big|_{initial}
\label{acc2} \\
& \Longrightarrow & {\rm for} \;\; t_{initial} \; \rightarrow \; 0
\, : \; R_{future} \; \gg \; R_{past} \;\; . \label{acc3}
\end{eqnarray}
\\ [4pt]
${\bullet \;}$ {\bf Single-Scalar Inflation:}
\\ [2pt]
Although the evidence for a phase of primordial inflation is very
strong \cite{Guth}, there is no compelling mechanism for making it
happen \cite{newinf}. The simplest model consists of gravity plus
a minimally coupled scalar field (called the {\it inflaton}) whose
Lagrangian is \cite{chaotic}:
\begin{equation}
{\cal L} \; = \;
\sqrt{-g} \left( -\frac{1}{2} \, g^{\mu\nu}
\partial_{\mu} \varphi \, \partial_{\nu} \varphi \; - \;
V(\varphi) \; + \;
\frac{R}{16 \pi G} \right)
\;\; . \label{Lscalar}
\end{equation}
Note that this model is general enough to support {\it any}
expansion history $a(t)$, provided only that $\dot{H}(t) \; \leq 0$
throughout. To see this, note that the nontrivial Einstein equations
are:
\begin{eqnarray}
3 H^2 &\!\!\! = \!\!\!&
8 \pi G \left[ \, \frac{1}{2} \Big( \frac{d\varphi}{dt} \Big)^2
+ V(\varphi) \right]
\;\; , \label{eom1scalar} \\
- 2\dot{H} + 3H^2 &\!\!\! = \!\!\!& 8 \pi G \left[ \, \frac{1}{2}
\Big( \frac{d\varphi}{dt} \Big)^2 - V(\varphi) \right] \;\;
.\label{eom2scalar}
\end{eqnarray}
One would usually take the scalar potential $V(\varphi)$ as given
and use these equations to determine the expansion history, but let
us adopt the opposite perspective. That is, we will assume $a(t)$ is
known and we then use the equations to reconstruct the potential
$V(\varphi)$ which supports that geometry. By adding (\ref{eom2scalar}) 
to (\ref{eom1scalar}) we get the inflaton as a function of time:
\begin{equation}
-2 \dot{H} \; = \;
8 \pi G \, \Big( \frac{d\varphi}{dt} \Big)^2
\quad \Longrightarrow \quad
\varphi(t) \; = \;
\varphi_I + \int^t dt' \left(
-\frac{\dot{H}(t')}{4 \pi G} \right)^{\frac{1}{2}}
\;\; . \label{varphi}
\end{equation}
By inverting this relation we get the time as a function of the
inflaton: $ t = t(\varphi) $. Now subtract (\ref{eom2scalar}) from
(\ref{eom1scalar}) to find the potential which gives the desired
expansion history:
\begin{equation}
6 H^2 \; = \;
16 \pi G \, V(\varphi)
\quad \Longrightarrow \quad
V(\varphi) \; = \;
\frac{3}{8 \pi G} \, H^2[t(\varphi)]
\;\; . \label{potential}
\end{equation}
This construction seems to have first appeared in \cite{TW1}, and
independently in \cite{SRSS} and \cite{CNO}.
\\ [4pt]
${\bullet \;}$ {\bf Scalar Inflation Problems:}
\\ [2pt]
As we have seen, the potential energy of a minimally coupled scalar
field can cause inflation, but this mechanism involves assumptions
which seem unlikely and are sometimes contradictory:
\\ [2pt]
{$\bullet \,$} That the universe began with the scalar field
approximately spatially homogeneous over more than a Hubble volume
$V(\varphi) > H^{-3}$ \cite{VT}.
\\ [2pt]
{$\bullet \,$} That the scalar field potential must be flat enough
make inflation last a long time \cite{newinf,chaotic}.
\\ [2pt]
{$\bullet \,$} That the minimum of the scalar field potential
has just the right value $V_{min} \simeq 0$ to leave the
post-inflationary universe with only the small amount of vacuum
energy we detect today \cite{SNIA,Yun}.
\\ [2pt]
{$\bullet \,$} That the scalar field couples enough to ordinary
matter so that its kinetic energy can create a hot, dense universe
at the end of inflation, but not so much that loop corrections from
ordinary matter compromise the flatness of the inflaton potential
\cite{Robert}.
\\ [4pt]
${\bullet \;}$ {\bf Gravity-Driven Inflation:}
\\ [2pt]
A more natural mechanism for inflation can be found within
gravitation -- which, after all, plays the dominant role in
shaping cosmological evolution -- by supposing that the bare
cosmological constant $\Lambda$ is not unnaturally small but
rather large and positive. Here ``large'' means a $\Lambda$
induced by some matter scale which might be as high as $10^{18} \,
GeV$. Then, the value of the dimensionless coupling constant
would be $\, G \Lambda \sim 10^{-4}$, rather than the putative 
value of $10^{-122}$ \cite{SNIA,Yun}.

Because $\Lambda$ is constant in {\it space}, no special
initial condition is needed to start inflation. We also
dispense with the need to employ a new, otherwise undetected
scalar field. However, $\Lambda$ is constant in {\it time}
as well, and classical physics can offer no natural mechanism
for stopping inflation once it has begun \cite{stability}.
Quantum physics can: accelerated expansion continually rips
virtual infrared gravitons out of the vacuum \cite{gravitons}
and these gravitons attract one another, thereby slowing
inflation \cite{NctRpw}. This is a very weak effect for
$\, G \Lambda \ll 1$, but a cumulative one, so inflation
would last a long time for no other reason than that gravity is
a weak interaction \cite{NctRpw}.
\\ [4pt]
${\bullet \;}$ {\bf Graviton Physical Modes:}
\\ [2pt]
In terms of the full metric field $g_{ij}(x) \,$,
the fluctuating graviton field $h_{ij}^{TT}(x)$ is
defined as:
\begin{equation}
g_{ij}(t, {\bf x}) \; = \;
a^2(t) \Big[ \delta_{ij} +
\sqrt{32 \pi G \,} \, h_{ij}^{TT}(t, {\bf x}) \Big]
\;\; . \label{hij}
\end{equation}
The free field expansion of the graviton field is:
\begin{equation}
h_{ij}^{TT}(t, {\bf x}) \; = \;
\int \frac{d^3 k}{(2 \pi)^3} \sum_{\lambda}
\left\{ u(t, k) \, e^{i {\bf k} \cdot {\bf x}}
\epsilon_{ij}({\bf k}, \lambda) \,
\alpha({\bf k}, \lambda) \, + \, (c.c.) \right\}
\;\; , \label{hijexp}
\end{equation}
where $(c.c.)$ denotes complex conjugation,
$\epsilon_{ij}({\bf k}, \lambda)$ are the same transverse
and traceless polarization tensors as in flat space,
$\alpha({\bf k}, \lambda)$  is the annihilation operator,
and $u(t, k)$ are the mode functions which obey:
\begin{equation}
{\ddot u}(t, k) \, + \,
3 H(t) \, {\dot u}(t, k) \, + \,
\frac{k^2}{a^2(t)} \, u(t, k)
\; = \; 0
\;\; . \label{modeseqn}
\end{equation}

The mechanism we have sketched is that inflation rips
gravitons out of the vacuum, and then the self-gravitation
of these particles slows inflation. Let us first estimate
the energy $E(t, k)$ which is present in a single polarization
of a single wave vector $\mathbf{k}$ at time $t$. Because the
precise definition of energy is subtle for gravitons we base
this estimate on a massless, minimally coupled scalar field
$\varphi(x)$, whose mode equation is the same as (\ref{modeseqn}).
The scalar field Lagrangian density is:
\begin{equation}
{\cal L}(x) \; = \;
- \frac12 \, \sqrt{-g} \, g^{\mu\nu} \,
{\partial}_{\mu} \varphi \,
{\partial}_{\nu} \varphi
\; = \;
\frac12 \, a^3(t) \, {\dot \varphi}^2 \, - \,
\frac12 \, \nabla \varphi \cdot \nabla \varphi
\;\; . \label{Lmmcs}
\end{equation}
The Langangian diagonalizes in momentum space:
\begin{equation}
L(t) =
\int d^3 x \; {\cal L}(x) =
\int \frac{d^3 k}{(2 \pi)^3}
\left\{ \frac12 \, a^3(t) \,
\Big\vert \dot{\widetilde{\varphi}}(t, {\bf k}) \Big\vert^2
- \, \frac12 \, a(t) \, k^2 \,
\Big\vert \widetilde{\varphi}(t, {\bf k}) \Big\vert^2
\right\}
\label{Ldiag}
\end{equation}
so that any mode with wavenumber ${\bf k}$ evolves independently as
a harmonic oscillator $q(t)$ with a time-dependent mass $m(t) \sim
a^3(t)$ and angular frequency $\omega(t) \equiv k \, a^{-1}(t)$:
\begin{eqnarray}
q(t) &\!\! = \!\!&
u(t, k) \, A \, + \, u^*(t, k) \, A^{\dagger}
\quad , \quad
\Big[ \, A \, , \, A^{\dagger} \, \Big] \, = \, 1
\;\; , \label{SHO1} \\
E(t,k) &\!\! = \!\!&
\frac12 \, a^3(t) \, {\dot q}^2(t) \, + \,
\frac12 \, a(t) \, k^2 \, q^2(t)
\;\; . \label{SHO2}
\end{eqnarray}
For the special case of de Sitter the mode functions are given by:
\begin{equation}
u(t,k) \, = \, \frac{H}{\sqrt{2 k^3}} \Biggl[ \,
1 - \frac{i k}{H \, a(t)}\, \Biggr] \;
\exp \! \left( \frac{ik}{H \, a(t)} \right)
\;\; . \label{mode&cc}
\end{equation}
Although our conclusions are quite generic, it will simplify the
subsequent analysis if we make this assumption of de Sitter.

At any instant $t$ the minimum energy is $E_{\rm min}(t, k)
= \frac12 k a^{-1}(t)$. However because both the mass
and angular frequency are time-dependent, the state with
minimum energy at one instant is not the minimum energy state
at later times; there is particle production. Bunch-Davies
vacuum $\vert \Omega \rangle$ is the minimum energy state in
the distant past, and the expectation value of the energy
operator (\ref{SHO2}) in this state is:
\begin{eqnarray}
\langle \Omega \vert \, E(t, k) \, \vert \Omega \rangle
\!\!& = &\!\!
\frac{a^3(t)}{2} \, \vert {\dot u}(t, k) \vert^2 \, + \,
\frac{k^2 a(t)}{2} \, \vert u(t, k) \vert^2 
\label{E1} \\
\!\!& = &\!\!
\frac{k}{a(t)} \Biggl( \, \frac12 + 
\Biggl[ \frac{H a(t)}{2 k}\Biggr]^2 \, \Biggr) 
\;\; . \label{E2}
\end{eqnarray}
By setting this equal to $(\frac12 + N) \, \hbar \omega$, one can 
read off the instantaneous occupation number $N(t,k)$:
\begin{equation}
N(t,k) \, = \, \Biggl[ \,
\frac{H \, a(t)}{2 k} \, \Biggr]^2
\;\; . \label{Ngrowth}
\end{equation}
We can consider $N(t,k)$ to be the number of gravitons with one
polarization and wave vector $\mathbf{k}$ that have been created
by time $t$.

At this point a short digression is useful on the significance of
the co-moving wave number $k$ in an expanding universe. Because $k =
2\pi/\lambda$ is the inverse of a coordinate length, the physical
wave number is $k a^{-1}(t)$. This falls exponentially during inflation.
Horizon crossing is when the physical wave number equals the Hubble
parameter:
\begin{equation}
{\it Horizon \; Crossing} \quad \Longrightarrow \quad
k_{\rm phys} \, = \, k \; a^{-1}(t)
\, = \, H
\;\; . \label{horcross}
\end{equation}
It is natural to separate modes into ``infrared'' and
``ultraviolet'' depending upon whether or not they have 
experienced horizon crossing:
\begin{eqnarray}
{\it Infrared} \quad & \Longrightarrow & \quad H < k < H \, a(t)
\;\; , \label{inflirmodes} \\
{\it Ultraviolet} \quad & \Longrightarrow & \quad k > H \, a(t) \;\;
. \label{influvmodes}
\end{eqnarray}
From (\ref{Ngrowth}) we see that there is negligible production of
ultraviolet gravitons, whereas the number of infrared gravitons in
even a single wave vector grows exponentially. This is a crucial
observation because it means that the physics of this effect is
controlled by the known, low energy theory of gravity, without
regard to its still unknown ultraviolet completion.

The energy density induced by both polarizations of these infrared
gravitons equals:
\begin{equation}
\rho_{IR} \; = \;
\frac{2}{a^3(t)} \int^{Ha} \; \frac{d^3k}{(2 \pi)^3} \;
N(t,k) \; \frac{k}{a(t)} \; = \;
\frac{H^4}{8 \pi^2}
\;\; . \label{rhoIR}
\end{equation}
This is much less than the energy density of the cosmological 
constant:
\begin{equation}
\rho_{\Lambda} \; = \;
\frac{3H^2}{8 \pi G}
\quad \Longrightarrow \quad
\frac{\rho_{IR}}{\rho_{\Lambda}} \; = \;
\frac{GH^2}{3 \pi} \; \ltwid \; 10^{-11}
\;\; . \label{rhoL}
\end{equation}
One may wonder if the gravitational self-interaction of $\rho_{IR}$
can even screen itself, much less $\rho_{\Lambda}$. To see that 
it can, note that even a small energy density can induce significant 
screening if it interacts over a {\it sufficiently large volume}. 
A simple way to see this is to consider the total energy density 
$\rho_{\rm tot}$ produced by a static energy density $\rho_{\rm bare}$ 
distributed throughout a sphere of radius $R$. For simplicity, we 
follow ADM \cite{ADM} in using the Newtonian formula assuming it is 
the total mass $\, \frac43 \pi \rho_{\rm tot} \, c^{-2} R^3 \,$ 
that gravitates:
\begin{equation}
\rho_{\rm tot} \approx
\rho_{\rm bare} -
\frac{4 \pi G \rho^2_{\rm tot} R^2}{5 c^4}
\Longrightarrow
\rho_{\rm tot} \approx
\frac{5 c^4}{8\pi G R^2} \left[
\sqrt{1 + \frac{16 \pi G \rho_{\rm bare} R^2}{5 c^4}}
- 1 \right]
\label{static}
\end{equation}
As $R$ goes to infinity the screening becomes total -- i.e.,
$\rho_{\rm tot}$ goes to zero -- independent of how small
$\rho_{\rm bare}$ is.

Equation (\ref{static}) means the gravitational self-interaction of
infrared gravitons can screen $\rho_{IR}$, but what about the vastly
larger energy density $\rho_{\Lambda}$ of the cosmological constant?
The key observation for realizing that even $\rho_{\Lambda}$ can be
screened is that {\it the gravitational self-interaction hasn't had
time to reach a static limit}. Indeed, most of the universe is not
even now in causal contact, and never will be if the current phase
of accelerated expansion persists. The lower bound of $\rho_{tot} =
0$ implicit in the static result (\ref{static}) arises because it is
the instantaneous value of $\rho_{tot}$ which gravitates, so making
it smaller by screening also cuts off the effect. But that cutoff
disappears when one takes account of the causal nature of the
interaction. The source for the gravitational field at time $t$ is
not the instantaneous energy density of infrared gravitons but
rather its value far back in the past light-cone. That is not
reduced by the instantaneous energy density becoming small; indeed,
the effect of screening is to make the past light-cone open
outwards, which exposes more of the early times when the energy
density of infrared gravitons was high.

This discussion does not prove the viability of gravity-driven
inflation. Because inflationary particle production is itself a 
1-loop effect, the gravitational response to it cannot occur at 
less than 2-loop order. Two-loop computations in quantum gravity are
not simple around flat space background, and they are considerably
tougher around de Sitter. Then there is the delicate gauge issue of
how to invariantly quantify screening \cite{TW2}. Good physicists on
both sides of the question have debated whether or not there is a
significant screening effect from the mechanism we have described
\cite{procon}, or from any of the related relaxation mechanisms
which have been proposed \cite{oldclaims,Polyakov,destab}. There is
even disagreement about the basic formalism of perturbative quantum
gravity on de Sitter background \cite{props,debate,weyl}. The aim of
this introduction has been merely to establish the {\it plausibility} 
of the mechanism. Having hopefully done that, we will henceforth explore 
a simple class of effective field equations that might describe it.

\section{Model Building}

${\bullet \;}$ {\bf Perturbative Results:}
\\ [2pt]
Let use first review some perturbative results on de Sitter:
\begin{equation}
{\rm de \; Sitter \; Inflation}
\quad \Longrightarrow \quad
a(t) \, = \, e^{H t}
\;\; . \label{dS}
\end{equation}
The gravitational Lagranian is:
\begin{equation}
{\cal L}_{gr} \; = \;
\frac{1}{16 \pi G} \big( R - 2 \Lambda \Big) \sqrt{-g}
\;\; . \label{Lgr}
\end{equation}
It turns out that quantum corrections cannot grow faster than powers
of $\ln(a) = Ht$ \cite{stochastic}. We are interested in the regime 
of $\ln(a) \gg 1$, in which case one needs only the {\it leading 
logarithm} contributions at any loop order $L$ which contain the most 
factors of $\ln(a)$. Explicit computations \cite{NctRpw,TW3}, and 
general counting rules \cite{stochastic}, give the following behaviour 
for the leading logarithm contributions to the energy density induced 
by quantum gravitational effects:
\begin{eqnarray}
\rho_1 &\!\! \sim \!\!& + \Lambda^2
\;\; , \\
\rho_2 &\!\! \sim \!\!& -G\Lambda^3 \, {\rm ln}[a(t)]
\;\; , \\
\rho_L &\!\! \sim \!\!& -\Lambda^2
\Big( G \Lambda \, {\rm ln}[a(t)] \Big)^{L-1}
\;\; . \label{pertresults}
\end{eqnarray}
Because stress-energy is separately conserved at each loop order,
the quantum gravitationally induced pressure must be that of
negative vacuum energy, up to small subleading logarithm
corrections:
\begin{equation}
\dot{\rho}_L \; = \; -3 H (\rho_L + p_L) 
\quad \Longrightarrow \quad
p_L(t) \sim -\rho_L(t) \;\; . \label{conserve}
\end{equation}
Hence the general form of the pressure is:
\begin{equation}
p(t) \; \sim \;
\Lambda^2 f[G \Lambda \, {\rm ln}(a)]
\;\; . \label{pressure}
\end{equation}

Perturbation theory is valid only if the effective dimensionless
coupling constant $G \Lambda \, {\rm ln}(a)$ of the theory is small.
Thus, perturbation theory breaks down after a large number of
e-foldings -- $N \equiv H \, t = {\rm ln}(a) \sim (G \Lambda)^{-1}$.
However, if we had the effective field equations, at least for a
general $FRW$ geometry, it would be possible to evolve arbitrarily 
far in the future. So we shall try to guess these equations based 
on some general principles, and on what we know from perturbation
theory. 
\\ [4pt] 
${\bullet \;}$ {\bf Guessing the Effective Field Equations:}
\\ [2pt]
The classical gravitational equations of motion coming from
(\ref{Lgr}) are:
\begin{equation}
G_{\mu\nu} \; = \;
- \Lambda g_{\mu\nu} \;\; . \label{eom1}
\end{equation}
The equations of motion in the presence of the quantum induced
stress-energy tensor $T_{\mu\nu}[g]$ are:
\begin{equation}
G_{\mu\nu} \; = \;
- \Lambda g_{\mu\nu} + 8 \pi G \, T_{\mu\nu}[g]
\;\;. \label{eom2}
\end{equation}
The full quantum induced stress-energy encodes all information about
quantum gravity. For example, variations of it about flat space --
with $\Lambda = 0$ -- give all scattering amplitudes to all orders 
in perturbation theory. There is absolutely no chance we can guess
this, nor is there any need to do so. We require only the most
cosmologically significant part of the full effective quantum
gravitational equations; that is, a functional of the $FRW$ scale
factor $a(t)$.

A few basic principles can be used to guide us \cite{NctRpw2}:
\\ [2pt]
{\bf (i) Correspondence:} The form of $T_{\mu\nu}[g]$ must of course
reproduce the known results from perturbation theory about de Sitter
space.
\\ [2pt]
{\bf (ii) Non-locality:} It is easy to show that a purely local
$T_{\mu\nu}[g](x)$ can only lead to a constant change in the
cosmological constant. Note first that such a local 
$T_{\mu\nu}[g](x)$ must be composed of the Riemann tensor and its
derivatives. Now consider the de Sitter geometry for an arbitrary
Hubble parameter $H'$, not necessarily equal to the one associated
with $\Lambda = 3 H^2$. The Riemann tensor for this geometry reduces
to a constant times sums of products of the metric, and any
covariant derivative of it therefore vanishes:
\begin{equation}
R_{\rho\sigma\mu\nu} \; = \; {H'}^2 \, [ \, g_{\rho\mu} \,
g_{\sigma\nu} - g_{\rho\nu} \, g_{\sigma\mu} ] 
\qquad \Longrightarrow \qquad 
D_{\alpha} R_{\rho\sigma\mu\nu} = 0 
\;\; . \label{riemann}
\end{equation}
Hence any local stress-energy must reduce, for this geometry, to 
$\# {H'}^4 g_{\mu\nu}$, and the effective field equation would 
become:
\begin{eqnarray} G_{\mu\nu} \, = \,
-3H^2 g_{\mu\nu} + \# 8 \pi G {H'}^4 g_{\mu\nu} 
\, = \,
-3{H'}^2 \Biggl( \frac{H^2}{{H'}^2} -
\frac83 \pi G {H'}^2 \Biggr) g_{\mu\nu} 
\;\; . \label{Gmunu2}
\end{eqnarray}
This amounts to merely a renormalization of $\Lambda$:
\begin{equation}
\Lambda ' \; = \; \frac{9}{16 \pi G} \Biggl[\sqrt{1 + \frac{32}9 \pi
G \Lambda} - 1 \Biggr] \;\; . 
\label{Lambda'}
\end{equation}
If one began in this geometry -- which our actual renormalization
condition would require -- then there would never be any deviation
form it. We conclude that screening requires a non-local 
$T_{\mu\nu}[g]$.
\\ [2pt]
{\bf (iii) Causality:} The quantum induced stress-energy must be
both conserved and causal, in the sense that $T_{\mu\nu}[g](x)$
depends only upon metrics on or within the past light-cone of the
point $x^{\mu}$. We would normally ensure conservation by defining
the stress-energy from the variation of an invariant effective
action:
\begin{equation}
T_{\mu\nu}[g](x) \; = \; 
-\frac{2}{\sqrt{-g}} \cdot 
\frac{\delta \Gamma[g]}{\delta g^{\mu\nu}(x)} 
\qquad \Longrightarrow \qquad
D^{\nu} T_{\mu\nu} = 0 
\;\; . \label{Tmunu}
\end{equation}
However, this procedure conflicts with causality for the sort of
non-local contributions of greatest interest to us.

To understand the problem, consider the action of a point particle
$q(t)$. Suppose the action contains a non-local term of the form
$q(s) \times q(s - \Delta t)$. One might think that its non-locality
is safely confined to the past of $q(s)$, but any variation must
also affect the term $q(s - \Delta t)$. This gives rise to an
equation which depends on the future as well as the past:
\begin{eqnarray}
\Gamma[q] &\!\! = \!\!&
\int ds \; q(s) \, q(s-\Delta t)
\quad \Longrightarrow \\
\frac{\delta \Gamma}{\delta q(t)} &\!\! = \!\!&
\int ds \Big[ \delta(s-t) \, q(s-\Delta t) +
q(s) \, \delta(s-\Delta t -t) \Big] \\
&\!\! = \!\!&
q(t-\Delta t) + q(t+\Delta t)
\;\; . \label{qvariable}
\end{eqnarray}
This same problem must afflict any variation such as (\ref{Tmunu})
which is based on a non-local effective action that contains only a
single field.

The proper way to derive non-local effective field equations which
are both causal and conserved is by varying the Schwinger-Kedysh
effective action \cite{SK}. This avoids the single field conundrum
by employing two fields $g_{\mu\nu}^{\pm}$; with the $+$ sign
corresponding to the background metric during forward evolution and
the $-$ sign to backwards evolution. The stress-energy tensor of the
Schwinger-Keldysh formalism is the variation with respect to either
field, after which the two fields are set equal:
\begin{equation}
T_{\mu\nu}[g](x) \; = \; 
- \frac{2}{\sqrt{-g}} \cdot \frac{\delta \Gamma[g^+,g^-]}
{\delta g^+_{\mu\nu}(x)} \Bigg|_{g^\pm \; = \; g}
\;\; . \label{causalTmunu}
\end{equation}
One can show that the $+$ and $-$ contributions from fields at any
point ${x'}^{\mu}$ exactly cancel unless ${x'}^{\mu}$ is on or
within the past light-cone of $x^{\mu}$.

The Schwinger-Keldysh effective action is what one should use to
{\it derive} the correct effective field equations. However,
deriving anything is tough in quantum gravity. The point of this
exercise was to try {\it guessing} the most cosmologically
significant part of the effective field equations. Because it is
those equations we seek, not the effective action, we shall adopt
the shortcut of simply making an appropriately non-local and causal
ansatz for them, and then enforce conservation directly.
\\ [4pt]
${\bullet \;}$ {\bf Perfect Fluid Ansatz:}
\\ [2pt]
The ansatz must apply to all $FRW$ cosmologies. The ``perfect fluid''
form of $T_{\mu\nu}$ can represent any cosmology and in addition
provides enough free parameters to enforce conservation and
correspondence with perturbative results:
\begin{equation}
T_{\mu\nu}[g] \; = \;
(\rho + p) \, u_{\mu} \, u_{\nu} \, + \,
p \, g_{\mu\nu}
\;\; , \label{Tmn}
\end{equation}
Our stress-energy is defined by specifying three things: \\
{\it (i)} the energy density $\rho$ as a functional
of the metric tensor $\rho[g](x)$, \\
{\it(ii)} the pressure $p$ as a functional of the
metric tensor $p[g](x)$, and \\
{\it (iii)} the 4-velocity field $u_{\mu}$ as a
functional of the metric tensor $u_{\mu}[g](x)$,
chosen to be timelike and normalized:
\begin{equation}
g^{\mu\nu} \, u_{\mu} u_{\nu} = -1
\qquad \Longrightarrow \qquad
u^{\mu} \, u_{\mu ; \nu} = 0
\;\; .
\label{u}
\end{equation}
Because of the normalization (\ref{u}), only three of the components
of $u_{\mu}$ are algebraically independent. Hence our ansatz consists 
of five independent functionals in total. Stress-energy conservation:
\begin{equation}
D^{\mu} \, T_{\mu\nu} \; = \; 0
\;\; , \label{cons1}
\end{equation}
provides four equations and allows us to determine any four of 
these functionals in terms of the fifth. It turns out to be most
convenient to specify the induced pressure functional $p[g]$ and
then use conservation to obtain the form of the induced energy
density $\rho[g]$ and the 4-velocity $u_{\mu}[g]$, up to their
initial value data.
\\ [4pt]
${\bullet \;}$ {\bf Building p[g] :}
\\ [2pt]
We want the pressure $p[g](x)$ to be a causal, non-local functional
of the metric which reduces to the form (\ref{pressure}) in the de
Sitter limit. A very simple ansatz along these lines is:
\begin{equation}
p[g](x) \; = \; 
\Lambda^2 \, f \Bigl( -G \Lambda \, X[g](x) \Bigr)
\;\; , \label{inducedp}
\end{equation}
where $-X[g](x)$ is a dimensionless, non-local functional of the
metric that grows like $\ln(a)$ when the metric is de Sitter. A
natural way of incorporating causal non-locality is through the
inverse of some differential operator. The simplest choice for this
operator is the covariant scalar d'Alembertian:
\begin{equation}
\square \, \equiv \, \frac{1}{\sqrt{-g}} \;
\partial_{\mu} \Big( \,
g^{\mu\nu} \sqrt{-g} \; \partial_{\nu} \, \Big)
\;\; . \label{box}
\end{equation}
To make $X[g](x)$ dimensionless, we need to act the inverse of
$\square$ on a curvature scalar, the simplest choice for which is
the Ricci scalar $R$. We are therefore led to consider $X[g] =
\square^{-1} R$, with the inverse defined using retarded boundary
conditions.

\begin{figure}
\centerline{\epsfig{file=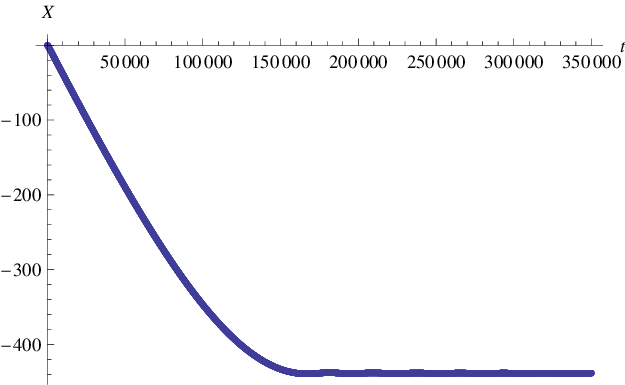,height=2.9in}}
\caption{\footnotesize The evolution of the source $X(t)$ over the
full range for the exponential model.}

\label{Xfull}

\end{figure}

\begin{figure}
\centerline{\epsfig{file=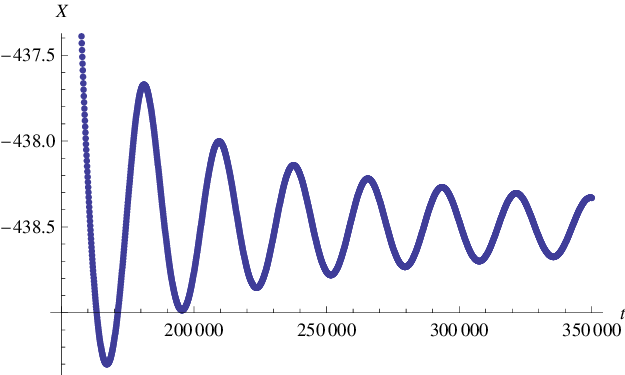,height=2.9in}}
\caption{\footnotesize The evolution of the source $X(t)$ during the
oscillatory regime for the \break \mbox{} \hspace{2cm} exponential
model.}

\label{Xosc}

\end{figure}

\begin{figure}
\centerline{\epsfig{file=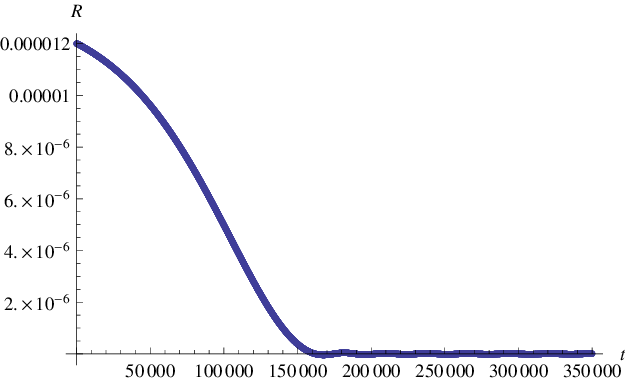,height=2.9in}}
\caption{\footnotesize The evolution of the curvature
scalar $R(t)$ over the full range for the
\break \mbox{} \hspace{2cm}
exponential model.}

\label{Rfull}

\end{figure}

\begin{figure}
\centerline{\epsfig{file=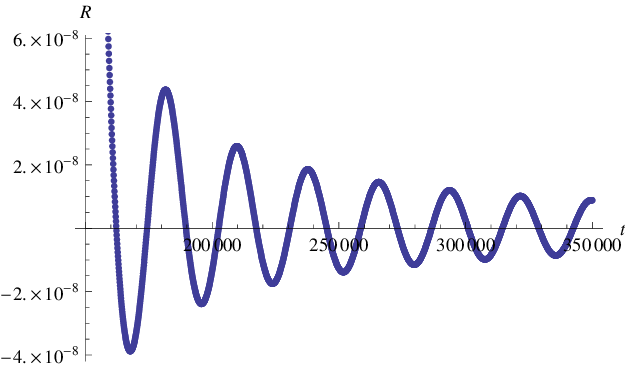,height=2.9in}}
\caption{\footnotesize The evolution of the curvature
scalar $R(t)$ during the oscillatory regime
\break \mbox{} \hspace{1.9cm}
for the exponential model.}

\label{Rosc}

\end{figure}

\begin{figure}
\centerline{\epsfig{file=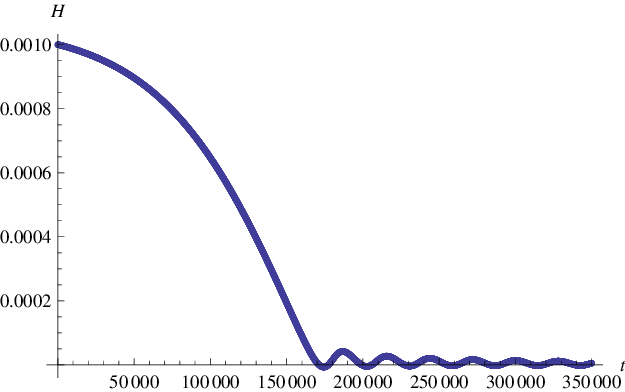,height=2.9in}}
\caption{\footnotesize The evolution of the Hubble
parameter $H(t)$ over the full range for the
\break \mbox{} \hspace{1.9cm}
exponential model.}

\label{Hfulleps}

\end{figure}

\begin{figure}
\centerline{\epsfig{file=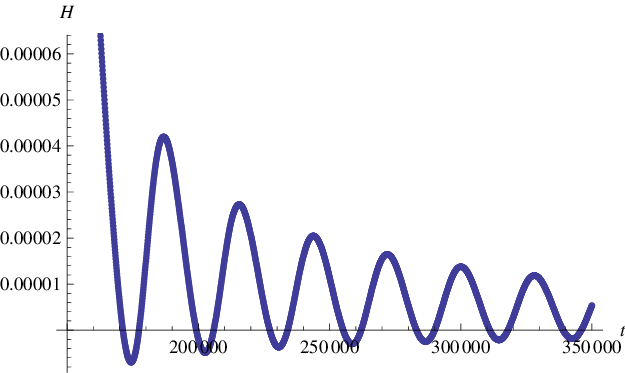,height=2.9in}}
\caption{\footnotesize The evolution of the Hubble
parameter $H(t)$ during the oscillatory regime
\break \mbox{} \hspace{1.8cm}
for the exponential model.}

\label{Hosceps}

\end{figure}

To see that this simple ansatz has the right properties, we
specialize $\square$ and $R$ to a general $FRW$ geometry:
\begin{equation}
\square \, = \, 
- \left( \, \partial_t^2 \, + \, 3H \partial_t \, \right) 
\qquad , \qquad 
R(t) \, = \, 
12 H^2(t) + 6\dot{H}(t) 
\;\; . \label{boxriccifrw}
\end{equation}
Hence the specialization of $X[g](x)$ to $FRW$ is:
\begin{equation}
X \; = \; 
\frac{1}{\square}R \; = \; 
- \int_0^t dt' \; a^{-3} \int_0^{t'} dt'' \; a^3 \, 
\Bigl[ 12 H^2 + 6\dot{H} \Bigr] 
\;\; . \label{Xfrw}
\end{equation}
For de Sitter spacetime -- $a(t) = e^{H t}$ with constant $H$
-- we get the correct correspondence limit:
\begin{equation}
\frac{1}{\square}R \; = \; 
-4 {\rm ln}(a) + \frac43 \left[ \, 1 - e^{-3 H t} \, \right] 
\;\; . \label{XdS}
\end{equation}
More generally, expression (\ref{Xfrw}) implies that $-X[g](x)$ will
grow during the inflationary regime of large Ricci curvature, and
then freeze in to a constant during the radiation dominated era of
$R(t) = 0$. As long as the function $f(x)$ in (\ref{inducedp}) grows
monotonically and without bound, this ansatz for the pressure is
bound to produce enough screening to end inflation in roughly the
right way.
\\ [4pt]
${\bullet \;}$ {\bf Numerical Results:}
\\ [2pt]
There is no hope of deriving an analytic solution for $a(t)$ when
the pressure is as complicated as (\ref{inducedp}) with
(\ref{Xfrw}), but this is a simple problem to solve numerically.
Figures \ref{Xfull}-\ref{Xosc} give the evolution of the non-local
source $X(t)$, figures \ref{Rfull}-\ref{Rosc} present the Ricci
scalar $R(t)$, and figures \ref{Hfulleps}-\ref{Hosceps} show the
Hubble parameter $H(t)$. These results were generated for the choice
$f(x) = \exp(x) - 1$ -- the ``exponential model'' -- although any
function $f(x)$ which grows monotonically and without bound gives
the same qualitative behaviour, including even $f(x) = x$. To avoid 
a long preliminary evolution with negligible effect, we set the
unrealistically high value of $G\Lambda = 1/200$. Again, the
behaviour is qualitatively the same for any choice of $G \Lambda$.

The following basic features emerge from our numerical work
\cite{NctRpw2}:
\\ [3pt]
-- During the era of inflation, the source $-X(t)$ grows while the
curvature scalar $R(t)$ and the Hubble parameter $H(t)$ decrease.
\\ [3pt]
-- Inflationary evolution dominates roughly until we reach the
critical point $X_{cr}$ defined by:
\begin{equation}
1 - 8 \pi G \Lambda \, f[ - G \Lambda \, X_{cr}]
\; \equiv \; 0
\;\; . \label{Xcr}
\end{equation}
-- The epoch of inflation ($q < 0$) ends slightly before $X(t)$
reaches $X_{cr}$. This is most directly seen from the deceleration
parameter because initially $q(t=0) = -1$, while at criticality
$q(t=t_{cr}) = +\frac12$.
\\ [3pt]
-- The source $X(t)$ oscillates with constant period and decreasing
amplitude.
\\ [3pt]
-- Oscillations in $R(t)$ become significant as we approach the end
of inflation; they are centered around $R = 0$, their frequency is
given by:
\begin{equation}
\omega \; = \;
G \Lambda H_0 \sqrt{72 \pi \, f_{cr}'}
\;\; , \label{omega}
\end{equation}
and their amplitude decreases like the inverse of the number of 
oscillations.
\\ [3pt]
-- While there is net expansion during the era of oscillations, the
Hubble parameter $H(t)$ attains small negative values for short time
intervals. Of course negative $H(t)$ corresponds to a compressing
universe, which should lead to rapid reheating when matter couplings
are included.

\newpage

${\bullet \;}$ {\bf Analytic Results:}
\\[2pt]
Although one cannot obtain analytic results for the full evolution
of $a(t)$, it is possible to give an approximate treatment for the
period of oscillations. We use the evolution equation:
\begin{equation}
2 {\dot H} + 3 H^2 \; = \; 
\Lambda \, \Big\{ \, 1 \, - \, 8 \pi G\Lambda \, 
f[ -G\Lambda \, X ] \, \Big\} 
\qquad , \qquad 
X \; \equiv \; \frac{1}{\square} \, R 
\;\; . \label{evoeqn1}
\end{equation}
Recall that we assume only that the function $f(x)$ grows
monotonically and without bound. Hence there must exist a critical
point $X_{cr}$ such that:
\begin{equation}
1 \, - \, 8 \pi G \Lambda \, f[ -G\Lambda \, X_{cr} ] 
\; = \; 0
\quad \Longrightarrow \quad 
X_{cr} \; = \; - \, \frac{1}{G\Lambda} \; 
f^{-1} \Big( \frac{1}{8 \pi G\Lambda} \Big) 
\;\; . \label{Xcrit}
\end{equation}
Inflationary evolution dominates roughly until we reach the critical
point. Close to the critical point the induced pressure $p$ is
nearly constant and, thus, it makes sense to expand $f$ around its
critical point:
\begin{equation}
f  \; \simeq \; 
f_{cr} - G \Lambda \, \Delta X(t) \, f'_{cr} 
\quad , \quad 
\Delta X(t) \; \equiv \; X(t) - X_{cr} 
\;\; . \label{fexpand}
\end{equation}
Now consider the linearized evolution equation:
\begin{equation}
2 {\dot H} + 3 H^2 \; \simeq \; 
8 \pi (G \Lambda)^2 \, \Lambda \, (X
- X_{cr}) \, f'[ -G\Lambda \, X_{cr} ] 
\;\; . \label{evoeqn2}
\end{equation}
Using (\ref{boxriccifrw}) we can express the co-moving time
derivative of the Hubble parameter as:
\begin{equation}
{\dot H} \; = \;
\frac16 R \, - \, 2H^2
\;\; . \label{Hdot}
\end{equation}
Because the amplitudes of both $R(t)$ and $H(t)$ fall like $t^{-1}$
during the era of oscillations, the second term in (\ref{Hdot}) is
irrelevant. Consequently, the evolution equation (\ref{evoeqn2})
becomes:
\begin{equation}
-R \, + \, 3 H^2 \; \simeq \; 
-24 \pi \, (G \Lambda)^2 \, \Lambda (X - X_{cr}) \, f'_{cr} 
\;\; , \label{evoeqn3}
\end{equation}
where we have defined:
\begin{equation}
f'_{cr} \; \equiv \; 
f'[ -G \Lambda \, X_{cr} ] \; \equiv \; 
- \frac{1}{G \Lambda} \, \frac{d}{dX} f[ -G \Lambda \, X ]
\Big\vert_{X = X_{cr}} 
\;\; . \label{f'cr}
\end{equation}
Action of the d'Alembertian operator (\ref{boxriccifrw}) on
(\ref{evoeqn3}) gives:
\begin{equation}
\ddot{R} \, + \, 2H \, \dot{R} \, + \,
( \omega^2 - {\dot H} )\, R \, + \,
\Big[ \, 3H^2 R - 36H^4 \, \Big]
\; \simeq \; 0
\;\; , \label{evoeqn4}
\end{equation}
where we define:
\begin{equation}
\omega^2 \; \equiv \;
24 \pi \, (G\Lambda)^2 \Lambda \,
 f'_{cr}
\;\; , \label{Omega}
\end{equation}
We can again neglect the various ``small'' terms in (\ref{evoeqn4})
to infer:
\begin{equation}
\ddot{R} \, + \, 2H \, \dot{R} \, + \, \omega^2 \, R \simeq 0 
\qquad \Longrightarrow \qquad 
R(t) \; \simeq \; \frac{\sin(\omega t)}{a(t)}
\;\; . \label{solevoeqn4}
\end{equation}
This reveals the presence of oscillations. Note also that the
frequency (\ref{Omega}) agrees with numerical results.

${\bullet \;}$ {\bf Generic Results:}
\\[2pt]
It is also possible to derive approximate analytic results for the
period of inflation. If $N$ is the number of e-foldings before
criticality, the various geometrical parameters are
\cite{NctRpw3,romania}:
\begin{eqnarray}
a(t) &\!\! = \!\!&
a_{\rm cr} \, e^{-N}
\;\; , \label{ainfl} \\
H(t) &\!\! \simeq \!\!&
\frac13 \, \omega \, \sqrt{ 4N + \frac43}
\;\; , \label{Hinfl} \\
\epsilon(t) &\!\! \simeq \!\!& \frac{2}{4N + \frac43} \;\; .
\label{epsiloninfl}
\end{eqnarray}

During the oscillatory era it is best to describe these same
parameters using the time $\Delta t \equiv t - t_{cr}$ since
criticality. The following approximate relations hold
\cite{NctRpw3,romania}:
\begin{eqnarray}
a(t) &\!\! = \!\!&
a_{\rm cr} \, C_2 \Big[ \,
C_1 \, + \, \omega \, \Delta t \, + \,
{\sqrt 2} \, \cos( \omega \, \Delta t + \phi ) \Big]
\;\; , \label{aosc} \\
H(t) &\!\! = \!\!&
\frac{ \omega \Big[ \, 1 -
{\sqrt 2} \, \sin( \omega \, \Delta t + \phi ) \Big] }
{ C_1 \, + \, \omega \, \Delta t \, + \,
{\sqrt 2} \, \cos( \omega \, \Delta t + \phi ) }
\;\; , \label{Hosc} \\
\epsilon(t) &\!\! = \!\!& 1 \, + \, \frac{ {\sqrt 2} \, \cos( \omega
\, \Delta t + \phi ) \Big[ \, C_1 \, + \, \omega \, \Delta t \, + \,
{\sqrt 2} \, \cos( \omega \, \Delta t + \phi ) \Big] } { \Big[ \, 1
- {\sqrt 2} \, \sin( \omega \, \Delta t + \phi ) \Big]^2 } \;\; .
\qquad \label{epsilonosc}
\end{eqnarray}
The constants $\phi$, $C_1$ and $C_2$ in relations
(\ref{aosc}-\ref{epsilonosc}) are chosen to match the two epochs at
criticality ($N=0$ and $\Delta t = 0$):
\begin{eqnarray}
\phi &\!\! = \!\!&
\arcsin \! \left(
\frac {\sqrt{2} - \sqrt{2970}}{56} \right)
\; \approx \;
- \frac{\pi}{2}
\;\; , \label{phi} \\
C_1 &\!\! = \!\!&
\frac{\sqrt{27}}{2} -
\frac{\sqrt{27}}{2} \, \sin\phi -
\sqrt{2} \, \cos\phi
\; \approx \; 3
\;\; , \label{C1} \\
C_2 &\!\! = \!\!&
\frac{1}{C_1 + \sqrt{2} \, \cos\phi}
\; \approx \; \frac16
\;\; . \label{C2}
\end{eqnarray}

${\bullet \;}$ {\bf Primordial Density Perturbations:}
\\[2pt]
{\it (i) Scalar perturbations.} Initially ultraviolet modes in
scalar driven inflation oscillate and redshift, and then become
approximately constant around the time of horizon crossing
\cite{Slava}. The behaviour of scalar perturbations in this model
differs in two significant ways. First, initially ultraviolet
modes merely redshift, they do not oscillate. Like scalar driven
inflation, the modes of this model become approximately constant
around the time of horizon crossing. However, all super-horizon
modes in this model begin oscillating with the same frequency
$\omega$ at the end of inflation \cite{NctRpw4}. Because there are
so many of these super-horizon modes after a long period of
inflation, the fact that all of them start to oscillate at the end
of inflation should lead to very rapid reheating, without the need
to invoke anything other than the usual gravitational couplings to
matter. After the universe reaches radiation domination one can show
that the oscillations stop \cite{NctRpw}, which is consistent with
an approximately scale invariant power spectrum. What we cannot do
is to evaluate the normalization. That is fixed by canonical
quantization in scalar driven inflation, but we only have the
effective field equations for this model. Recall that the
combination of causality and non-locality means our effective field
equations cannot derive from a conventional action principle. 
\\ [2pt]
{\it (ii) Tensor perturbations.} The analysis of tensor
perturbations in this class of models is much simpler than that of
scalar perturbations \cite{romania}. The reason is that our perfect
fluid stress-energy has no effect on the tensor perturbations
$h_{ij}^{TT}$. Therefore the resulting power spectrum ${\Delta}^2_h$
has the usual form:
\begin{equation}
\Delta^2_{h}(k) \, \simeq \, 
\frac{16 G H^2(t_k)}{\pi} 
\;\; , \label{tensorpert}
\end{equation}
but with the expansion history peculiar to our model. There is
nothing unconventional about our expansion history
(\ref{ainfl}-\ref{epsiloninfl}) during the epoch of inflation, so
our prediction for the $B$ mode of polarization in the cosmic
microwave background is not distinct from that of scalar driven
inflation. The period for which our model differs is the phase of
oscillations (\ref{aosc}-\ref{epsilonosc}), during which the usual
Hubble ``friction'' term actually changes sign. Because the end of
inflation comes about 50 e-foldings after the horizon crossing of
the observable part of the cosmic microwave background, the
corresponding enhancement in the stochastic background of
gravitational radiation will be at the uncomfortably high frequency
of $f \sim 10^9 Hz $ \cite{romania}. No current gravity wave detector 
has sensitivity at this frequency but one has been proposed \cite{MHz}.

\section{Post-Inflationary Evolution}

We assume that energy flows from the gravitational to the matter
sector, leading to a radiation dominated universe at $\; t = t_r$.
Because our model is sourced by the Ricci scalar, which vanishes
during radiation domination, the quantum induced stress-energy
simply cancels the bare cosmological constant. There is no deviation
from conventional cosmology until the onset of matter domination at
$\; t = t_m$. By that time the scales are so much below those of
primordial inflation that only very small changes occur in $X(t)$,
and we can use first order perturbation theory to compute the total
pressure:
\begin{eqnarray}
p_{\rm tot} \!\!& \equiv &\!\! 
- \, \frac{\Lambda}{8\pi G} \; + \; p[g](x) 
\label{ptotal1} \\
\!\!& = &\!\!
- \, \frac{\Lambda}{8\pi G} \, \Biggl\{
1 - 8 \pi G \Lambda \;
f\Big[ -G \Lambda \, (X_{cr} + \Delta X) \, \Big] \Biggr\}
\label{ptotal2} \\
\!\!& \simeq &\!\!
- \, \frac{\Lambda}{G}
\times (G \Lambda)^2 \, f_{cr}' \; \Delta X
\;\; . \label{ptotal3}
\end{eqnarray}
The simple source (\ref{inducedp}) grows according to the formula:
\begin{eqnarray}
\Delta X(t) \!\!& \equiv &\!\!
X(t) - X_{cr} \; = \;
- \frac43 \ln \Bigl[ \,
1 + \frac32 H_m (t - t_m) \, \Bigr]
\; + \; O(1)
\;\; . \label{DeltaX}
\end{eqnarray}
These facts give rise to two fatal problems for the model: 
\\ [2pt]
{\it - The Sign problem:} Because $f$ is monotonically increasing 
and unbounded:
\begin{equation}
p_{\rm tot} > 0 
\quad {\rm when} \quad 
X(t) < X_{cr} \ll 0 
\;\; . \label{sign}
\end{equation}
The observation of late time acceleration \cite{SNIA,Yun}
implies the opposite. 
\\ [2pt]
{\it - The Magnitude problem:} The magnitude of the total pressure
produced is unacceptably large:
\begin{eqnarray}
\frac{p_{\rm tot}}{p_{\rm now}}
\; \simeq \;
\left( \frac{G \Lambda \, H_I}{H_{\rm now}}
\right)^2 f_{cr}' \; \Delta X
\; \simeq \;
10^{86} \times f_{cr}' \times \Delta X
\;\; , \label{magn1}
\end{eqnarray}
where we have used:
\begin{equation}
p_{\rm now} \simeq 
- \, \frac{3}{8 \pi G} H_{\rm now}^2
\quad , \quad
H_I \sim 10^{13} GeV
\quad , \quad
H_{\rm now} \sim 10^{-33} eV
\;\; . \label{magn2}
\end{equation}
\\ [4pt]
${\bullet \;}$ {\bf Improved Ansatz:}
\\ [2pt]
Both problems can be addressed by changing the source (\ref{Xfrw}).
What we need to do is add an extra curvature $S$ inside the inverse
d'Alembertian, divided by $\Lambda$ to keep things dimensionless
\cite{NctRpw5}:
\begin{eqnarray}
p[g](x) &\!\! = \!\!&
\Lambda^2 \, f[- G \Lambda \, X](x)
\;\; , \\
- G \Lambda \, X &\!\! = \!\!&
- G \Lambda \; \frac{1}{\square} \, R
\quad \longrightarrow \quad
- G \; \frac{1}{\square}
\left( R \times S \, \right)
\; = \;
- \frac{G \Lambda}{\square}
\left( R \times \frac{S}{\Lambda} \, \right)
\qquad \label{magn3}
\end{eqnarray}
In this way the magnitude falls with cosmological evolution so that
$\Delta X(t)$ experiences only an acceptably small change at the
onset of matter domination. To keep inflation ending successfully it
is necessary to evaluate this curvature $S$ far back in the past of
the Ricci scalar $R$. We obtained acceptable results with a factor
of ten.

That suffices for the magnitude problem. To solve the sign problem
we note that the curvature scalar is positive during both inflation
-- $R = +12H^2$ -- and matter domination -- $R = +3H^2$. A simple
choice for $S$ that changes its sign is $R_{00}$ which equals
$-3H^2$ during inflation and $+\frac32 H^2$ during matter domination
\cite{NctRpw5}. Note that we can invariantly select the $00$
component of $R_{\mu\nu}$ using the timelike 4-velocity field
$u^{\mu}$, which is just $\delta^{\mu}_0$ for $FRW$. Hence the
specialization of the improved ansatz to $FRW$ is:
\begin{eqnarray}
p[g](x) &\!\! = \!\!&
\Lambda^2 \, f[- G \Lambda \, Y](x)
\;\; , \\
Y[g](t) &\!\! = \!\!&
- \frac1{\Lambda} \,
\frac{1}{\square} \left[ \,
R(t) \times R_{00}({\scriptstyle \frac{1}{10}} t)
\, \right]
\; \equiv \;
X_{cr} + \Delta Y
\;\; . \label{imprantsatz}
\end{eqnarray}

${\bullet \;}$ {\bf Late Time Acceleration:}
\\ [2pt]
Finally, we compute the total pressure in the improved ansatz
\cite{NctRpw5}:
\begin{eqnarray}
p_{\rm tot} \; \simeq \;
- \, G \Lambda^3 \, f_{cr}' \; \Delta Y
\; \simeq \;
- \, 200 \, G \Lambda^2 \, f_{cr}' \; H^2_m
\;\; . \label{imprptot1}
\end{eqnarray}
For the exponential model:
\begin{equation}
f(x) = e^x -1
\quad \Longrightarrow \quad
f_{cr}' = \frac{1}{8 \pi G \Lambda}
\;\; , \label{fexp}
\end{equation}
the pressure ratio is:
\begin{eqnarray}
t \gg t_m
\quad \Rightarrow \qquad
\frac{p_{\rm tot}}{p_{\rm now}}
\!\!& \simeq &\!\!
\frac{200}{3} \,
8 \pi (G \Lambda)^2 \times f_{cr}' \times
\left( \frac{H_m}{H_{\rm now}} \right)^2
\qquad \\
\!\!& \simeq &\!\!
\frac{200}{3} \,
8 \pi (G \Lambda)^2 \times f_{cr}' \times
10^{10}
\\
\!\!& \simeq &\!\!
\frac23 \times 10^{12} \times G \Lambda
\;\; . \label{imprptot2}
\end{eqnarray}
It is evident that for physically reasonable values of $\; G \Lambda
= M^4 \, M^{-4}_{Pl} \;$ we can achieve the desired equality of
$p_{\rm tot}$ with $p_{\rm now}$ whose ratio is given by
(\ref{imprptot2}).

\section{Conclusions}

There is very strong evidence that the universe underwent a very
early phase of accelerated expansion known as primordial inflation.
One can devise a scalar inflaton (\ref{Lscalar}) to support this
geometry but this entails positing a new and otherwise undetected
degree of freedom, as well as making some unrealistic and sometimes
contradictory assumptions about the inflaton's potential and its
initial condition. On the other hand, there is no question that
inflation results in the production of a vast sea of infrared
gravitons, nor is there any question that these gravitons attract
one another to some extent. Explicit results from perturbation
theory indicate that this attraction grows stronger with time, until
perturbation theory eventually breaks down.

Great controversy surrounds this final claim but, if it can be
established, the phenomenological payoff is enormous. For then it
becomes possible to dispense with the scalar inflaton and to make a
virtue out of what is usually regarded as a terrible problem:
namely, the fact that the observed cosmological constant is more
than 120 orders of magnitude below its natural scale. We propose
that the bare cosmological constant is not unnaturally small but
instead only a few orders of magnitude below the Planck scale. What
is being measured today is not this bare cosmological constant but
rather the expansion rate, and we propose that the effect of the
bare cosmological constant on the current expansion rate is subject
to almost perfect screening by the self-gravitation of gravitons
produced during a very long period of $\Lambda$-driven inflation.

We believe it is possible to use perturbation theory to establish
the reality of quantum gravitational screening. We also feel one can
resum the series of leading infrared logarithms to {\it derive} what
happens at late times. However, neither thing will be easy, nor will
they be quickly attained. In the meantime, we have devised a class
of non-local effective field equations which might describe the
eventual result of such a derivation. At this stage, one is free to
dismiss our motivation from quantum gravitational inflation and
simply regard these effective field equations in the same light as
another classical model of inflation. They are at least no worse
than scalar inflaton models, and they do have some remarkable and
quite generic features. Chief of these are that inflation ends in a
phase of oscillations which violate the weak energy condition, and
for which there is participation from every super-horizon mode, not
just the zero mode. The former feature may have left an observable
signature in the stochastic background of gravitational radiation
\cite{romania}. And the last feature should lead to almost
instantaneous reheating using only the universal gravitational
coupling to matter \cite{NctRpw4}.

Although the simplest of our models breaks down after the onset of
matter domination, it can be easily fixed. Indeed, this can be done
in such a way as to explain the current phase of cosmic acceleration. 
It will be interesting to see if any of these models can be derived 
from fundamental theory.

\vskip 1cm

\centerline{\bf Acknowledgements}

We are grateful to L. Papantonopoulos for the invitation to deliver
these lectures and for his extraordinary patience while they were
written up. This work was partially supported by the European Union
(European Social Fund, ESF) and Hellenic national funds through the
Operational Program ``Education and Lifelong Learning'' of the
National Strategic Reference Framework (NSRF) under the ``Thalis''
action MIS-375734, under the ``Funding of proposals that have
received a positive evaluation in the 3rd and 4th Call of ERC
Grant Schemes''; by NSF grant PHY-0855021, and by the Institute
for Fundamental Theory at the University of Florida.

\end{document}